\documentclass{PoS}

\usepackage{amsmath}

\newcommand{\onecol}[2]{
        \begin{minipage}[t]{#1}{#2\vfill} \end{minipage}
        }

\title{Fermions as global correction in lattice QCD}

\ShortTitle{Fermions as global correction in lattice QCD}

\author{
   \hfill
   \onecol{3cm}{\vspace{-2.5em}\it
      BUW-SC 2011/09\\
      WUB/11-22
   }
   \vspace{1cm}
}

\author{\speaker{Bj\"orn Leder}$^{a,b}$, Jacob Finkenrath$^a$ and
       Francesco Knechtli$^a$\\
       \llap{$^a$}Department of Physics, Bergische Universit\"at Wuppertal\\
                  Gaussstr. 20, D-42119 Wuppertal, Germany\\
       \llap{$^b$}Department of Mathematics, Bergische Universit\"at Wuppertal\\
                  Gaussstr. 20, D-42119 Wuppertal, Germany\\
       E-mail: \email{leder@physik.uni-wuppertal.de}
}

\abstract{The fermion determinant is a highly non-local object and its logarithm
 is an extensive quantity. For these reasons it is widely believed that the
 determinant cannot be treated in acceptance steps of gauge link configurations
 that differ in a large fraction of the links. However, for exact factorisations
 of the determinant that separate the ultraviolet from the infra-red modes of
 the Dirac operator it is known that the latter show less variation under changes
 of the gauge field compared to the former. Using a factorisation based on
 recursive domain decomposition allows for a hierarchical algorithm that starts
 with pure gauge updates of the links within the domains and ends after a number
 of filters with a global acceptance step. We find that the global acceptance rate
 is high on moderate lattice sizes. Whether this type of algorithm can help in curing
 the problem of critical slowing down is presently under study.}

\FullConference{ The XXIX International Symposium on Lattice Field Theory - Lattice 2011\\
July 10-16, 2011\\
Squaw Valley, Lake Tahoe, California}

\newcommand{\Pacc}{P_{\rm acc}}
\newcommand{\minm}[1]{\min\left\{1,#1\right\}}
\newcommand{\ev}[1]{\left\langle #1 \right\rangle}

\begin{document}

\section{Introduction}
\vspace{-1em}

In this contribution we analyse the acceptance rate of global acceptance steps
with the fermion determinant \cite{Knechtli:2003yt} in lattice QCD. The
type of algorithm detailed in the following sections is highly flexible with
respect to the fermion and gauge action used, because complicated and expensive
force computations are avoided. The proposed changes of the gauge configuration
come from a hierarchical filter (section \ref{sec:hierarchy}) that separates short and long distance physics.
Hierarchical filters based on approximations of the determinant with increasing accuracy
were introduced and tested in \cite{Hasenbusch:1998yb}.
The advantage of our approach, based on recursive domain decomposition, is that it allows
for a decoupling of updates within the domains and thus parallel domain-wise acceptance steps
in the filter (section \ref{sec:dd}). In \cite{Luscher:2003vf} a serial hierarchical
filter of increasing block size was proposed, but never tested.

The determinant in the global acceptance step has to be treated stochastically, but the
exact acceptance rate of the gauge link updates can nevertheless be determined
(section \ref{sec:stochastic}) and its volume dependence studied (section \ref{sec:exact}).
We demonstrate that for a lattice of size $(0.8\; \mathrm{fm})^4$ the exact
global acceptance rate is $\ge 60\%$  (section \ref{sec:run}).

\vspace{-1em}
\section{Hierarchical acceptance steps}
\label{sec:hierarchy}
\vspace{-1em}

Let $P(s)$ be the desired distribution of the states $s$ of a system. Suppose a
process that proposes a new state $s'$ with probability $P_0(s'\leftarrow s)$ and 
fulfils detailed balance with respect to $P_0(s)$. A process with fixed point distribution
$P(s)$ is then obtained by the iteration of a proposal with subsequent Metropolis
acceptance step \cite{metropolis:1087}
\begin{equation}
\begin{split}
 0) &\quad \text{Propose $s'$ according to $P_0(s'\leftarrow s)$}\\
 1) &\quad \Pacc(s'\leftarrow s) = \minm{\frac{P_0(s)P(s')}{P(s)P_0(s')}} \,.
\end{split}
\end{equation}
This hierarchy of a proposal step and an acceptance step, that in combination have
the correct fixed point distribution,
can easily be generalized to an arbitrary number of acceptance steps. The result
of the first acceptance step $1)$ is then interpreted as the proposal for a second
acceptance step $2)$ and so on. If the target distribution $P(s)$ factorises into
$n+1$ parts
\begin{equation}\label{eq:factor}
 P(s)=P_0(s)\,P_1(s)\,P_2(s)\dots P_{n}(s)\,,
\end{equation}
the resulting hierarchical acceptance steps take the form
\begin{equation}
\begin{split}
 0) &\quad \text{Propose $s'$ according to $P_0(s'\leftarrow s)$} \\
 1) &\quad \Pacc^{(1)}(s'\leftarrow s) = \minm{\frac{P_1(s')}{P_1(s)}} \\
 2) &\quad \Pacc^{(2)}(s'\leftarrow s) = \minm{\frac{P_2(s')}{P_2(s)}} \\
 ...\\
 n) &\quad \Pacc^{(n)}(s'\leftarrow s) = \minm{\frac{P_n(s')}{P_n(s)}} \,.
\end{split}
\end{equation}
In the context of lattice QCD it is plausible to assume $P_i(s)\propto \exp(-S_i(s))$
and thus $P_i(s')/P_i(s) = \exp(-\Delta_i(s',s))$ with $\Delta_i(s',s)=S_i(s')-S_i(s)$.
For Gaussian distributed $\Delta_i(s',s)$ with variance $\sigma_i^2$ the acceptance rate
in step $i)$ is \cite{Knechtli:2003yt}
\begin{equation}\label{eq:acceptance}
 \ev{\Pacc^{(i)}}_{s,s'} = {\rm erfc}\left(\sqrt{\sigma_i^2/8}\right)\,.
\end{equation}
The acceptance rates might be optimised by parametrising and tuning the factorisation
\eqref{eq:factor} \cite{algo}.

In 2-flavour lattice QCD a simple two-step algorithm would consist of some update
of the gauge link configuration $U$ according to the gauge action alone, e.g.
 $P_0(U)\propto \exp(-S_G(U))$, and an 
acceptance step with the fermion determinant
\begin{equation}
 \Pacc^{(1)}(U'\leftarrow U) = \minm{\det \frac{D'^\dagger D'}{D^\dagger D}} \,.
\end{equation}
If the proposal is a 1-link change and the Dirac operator $D$ involves only 
nearest neighbour couplings it is easy to show that the acceptance step requires
$\mathrm{O}(1)$ inversions. An ergodic algorithm is then obtained by sweeps through the lattice.
Thus the cost of such an algorithm would scale at least like $V^2$ \cite{Weingarten:1980hx}.

If, on the other hand, a finite fraction $\propto V$ of the links is updated for the proposal,
the distribution $P_1$ might be written as $P_1(U)\propto \exp(\ln(\det\, D^\dagger D))$.
Since the logarithm of the determinant of $D^\dagger D$ is an extensive quantity the variance
of the distribution of $\Delta_i(U',U)=\ln(\det\, D'^\dagger D') - \ln(\det\, D^\dagger D)$
is $\sigma_1^2\propto V$. From \eqref{eq:acceptance} one concludes that for $V\to\infty$
the acceptance rate decreases exponentially with the volume.

From the preceding discussion it is obvious that such two-step algorithms will not be efficient
for large lattices. Indeed numerical experiments show that for lattices larger than
$\sim(0.2\;\mathrm{fm})^4$ (where all links are updated) the acceptance rate
quickly becomes less than a percent. However, in the context
of low mode reweighting the fluctuations of the determinant of $D_{\text{low}}^\dagger D_{\text{low}}$,
where $D_{\text{low}}$ is a restriction of $D$ to its low modes, are found to
depend only mildly on the volume \cite{Luscher:2008tw}. The explanation for this observation
might be the fact that the fluctuations of the small eigenvalues of $D^\dagger D$
decrease like $1/V$ \cite{DelDebbio:2005qa}. Thus, given a factorisation of the determinant
that separates low (infra-red IR) and high (ultraviolet UV) modes
\begin{equation}
 \det(D) = \det(D_{\text{UV}}) \cdots \det(D_{\text{IR}})\,,
\end{equation}
a hierarchy of acceptance steps can be constructed, where the large fluctuations
of the UV modes go through a set of filters (acceptance steps) which are more and more dominated
by the IR modes:
\begin{center}
\begin{tabular}{cccccc}
$0)$ & $P_0$ & UV & short distance & local & cheap\\
& $\vdots$ & $\vdots$ & $\vdots$ & $\vdots$ & $\vdots$ \\
$n$) & $P_n$ & IR & long distance & global & expensive
\end{tabular}
\end{center}
This hierarchy of modes may induce also a hierarchy of costs since it is the
low modes that cause the most cost in lattice QCD. Furthermore the factorisation
should be exact and the terms simple to compute. Factorisations that realise these
conditions are already used to speed-up the HMC algorithm, i.e.~in the context of
mass-preconditioning \cite{Hasenbusch:2001ne} and domain-decomposition \cite{Luscher:2005rx}.
Only the latter also allows for a decoupling of local updates and will
be discussed in the following.

\vspace{-1em}
\section{Domain decomposition}
\label{sec:dd}
\vspace{-1em}

Domain decomposition was introduced in lattice QCD in \cite{Luscher:2003vf}
and in \cite{Luscher:2005rx} the
resulting factorisation of the fermion determinant was used to separate short
distance and long distance physics in the HMC algorithm.

\begin{figure}[tb]
 \centering
 \includegraphics[width=0.485\textwidth]{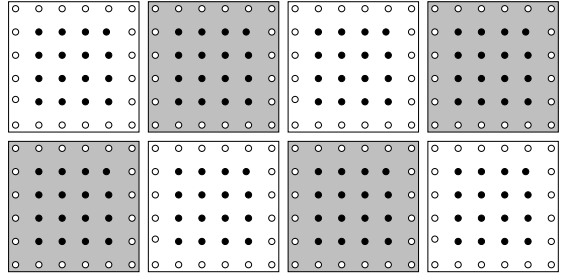}
 \caption{Block decomposition of a 2-dimensional lattice. The blocks are coloured like a checker board.
          Picture taken from \cite{Luscher:2005rx}}
 \label{fig:bgr}
\vspace{-1em}
\end{figure}

Suppose a decomposition $\mathcal{C}$ of the lattice in non-overlapping blocks $b\in \mathcal{C}$
(cf.~fig.~\ref{fig:bgr} for a 2-dimensional visualisation).
The lattice sites are labelled such that the
sites belonging to the first black block come first,
then the second black block and after the last black block the first white block
and so on. The Dirac operator can then be written in as
\begin{equation}\label{eq:domdec}
 D = \left(\begin{array}{cc} D_{\rm bb} & D_{\rm bw} \\
D_{\rm wb} & D_{\rm ww} \end{array}\right)\,,
\end{equation}
where $D_{\rm bb}$ ($D_{\rm ww}$) is a block-diagonal matrix with the black (white)
block Dirac operators $D_b$ on the diagonal. The block Dirac operators $D_b$ fulfil
Dirichlet boundary conditions and therefore are dominated by short distance physics
(if the blocks are small enough).
The matrices $D_{\rm bw}$ and $D_{\rm wb}$
contain the block interaction terms. The form \eqref{eq:domdec} induces a factorisation
of the determinant
\begin{equation}\label{eq:ddfactor}
 \det(D) = \prod_{b \in \mathcal{C}}\det(D_b)\det(\hat{D})\,, \quad \hat{D} = 1 - D_{\rm bb}^{-1}D_{\rm bw}D_{\rm ww}^{-1}D_{\rm wb}\,,
\end{equation}
where $\hat{D}$
is the Schur complement of the decomposition \eqref{eq:domdec} and contains block
interactions, i.e.~the long distance physics. A natural separation scale 
is given by the inverse block size $1/L_b$.
In the context of the domain decompositioned HMC the average force associated
with the Schur complement is an order of magnitude smaller than the
force associated with the block Dirac operators \cite{Luscher:2005rx}.
This indicates that the fluctuations of the determinant of the Schur complement are smaller
than that of the block determinants. Furthermore the factorisation \eqref{eq:ddfactor}
can be iterated using a recursive domain decomposition
\begin{equation}\label{eq:ddrec}
 \det(D_b) = \prod_{b' \in \mathcal{C}_b}\det(D_{b'})\det(\hat{D}_b)\,.
\end{equation}
In the case of one level of recursion the hierarchy of acceptance steps is given by 
\begin{equation}\label{eq:4step}
\begin{split}
 0) &\quad \text{Update \emph{active} links on all blocks (e.g. heat-bath/over-relaxation)} \\
 1) &\quad \Pacc^{(1)}(b,b') = \minm{\det \frac{D_{b'}'^\dagger D_{b'}'}{D_{b'}^\dagger D_{b'}}}\,,\quad \forall b, \forall b' \in \mathcal{C}_b \\
 2) &\quad \Pacc^{(1)}(b) = \minm{\det \frac{\hat{D}_b'^\dagger \hat{D}_b'}{\hat{D}_b^\dagger \hat{D}_b}}\,,\quad \forall b \in \mathcal{C} \\
 3) &\quad \Pacc^{(n)}({\rm global}) = \minm{\det \frac{\hat{D}'^\dagger \hat{D}'}{\hat{D}^\dagger \hat{D}}} \,.
\end{split}
\end{equation}

The set of \emph{active} links that is updated in step $0)$ is chosen such that
the block acceptance steps decouple. In the case of Wilson fermions with clover term
these are the links that have at most one endpoint on the boundary of a block (white
points in fig.~\ref{fig:bgr}). If the smallest blocks, $b'$, consist of not more than
$\sim 6^4$ lattice points, the determinant ratios in step $1)$ can be efficiently
computed exactly by sparse LU-decomposition. The Schur complements of steps $2)$ and
$3)$ are usually too large to be computed exactly and thus have to be treated
stochastically.

\newcommand{\varlnla}{\sigma^2_{\ln\lambda}}

\vspace{-1em}
\section{Stochastic acceptance step}
\label{sec:stochastic}
\vspace{-1em}

The determinants of ratios of block and the global Schur complements in \eqref{eq:4step}
are replaced by stochastic estimators. Dropping the subscripts and setting
$M = \hat{D}'^{-1}\hat{D}$ we replace the exact acceptance step by a stochastic acceptance
step
\begin{equation}
\minm{\det(M^\dagger M)} \to \minm{{\rm e}^{-\eta^\dagger(M^\dagger M-1)\eta}}\,,
\end{equation}
where $\eta$ is a complex Gaussian noise vector that is updated before each acceptance
step. Such an algorithm can be shown to fulfil detailed balance and yield an acceptance
rate that is bounded from above by the one of the exact acceptance step
\cite{Knechtli:2003yt,algo}.

Stochastic acceptance steps will only work if the variance of
the estimator can be controlled (c.f.~eq.~\eqref{eq:acceptance}). The eigenvalues
$\lambda$ of $M^\dagger M$  are real and the bulk of them follows a log-normal
distribution with mean zero and variance $\varlnla$.
It is easy to see that the estimator variance is not defined if the smallest
eigenvalues of $M^\dagger M$
is $\le 0.5$ \cite{Hasenfratz:2002ym} and that eigenvalues $\lambda=1$
do not contribute to it. Using a simple model one can further show that
the estimator variance is proportional to the number $\hat{N}_{1}$ of eigenvalues that are
not one and the variance $\varlnla$ of the log-normal distribution of the
eigenvalues \cite{diplom,algo}.

The fact that we are estimating the determinant ratio of the Schur complement and
not the full Dirac operator helps, since it acts non-trivially only on a subset of
the boundary points (white points in fig.~\ref{fig:bgr}) \cite{Luscher:2005rx},
i.e.~it reduces $\hat{N}_{1}$. The variance $\varlnla$ of the eigenvalue
distribution can be reduced by relative gauge fixing the old and the proposed
link configuration \cite{Knechtli:2003yt}. To ensure $\lambda>0.5$
and to balance cost and acceptance rate these measures turn out not to be sufficient.
Therefore a parametrisation of the proposed change in the gauge field is introduced
by a sequence $U_i,\;i=0,\ldots,N$ with $U_0=U$, $U_N=U^\prime$, inducing a factorisation
\begin{equation}\label{eq:param}
\det(M^\dagger M) = \prod_{i=0}^{N-1} \det(M_i^\dagger M_i)\,.
\end{equation}
Each factor is then replaced by a stochastic estimator with an independent noise vector.
The cost is then one inversion for each factor. We observe that
if the sequence of link configurations fulfils 
$||U_i- U_{i+1}||\propto 1/N \,,\; \forall\, i<N$ the variance $\varlnla$
and thus the variance of the estimator of this factor is reduced by $1/N^2$.\footnote{For 
this statement to hold, the plaquette should not change too much along the sequence
of link configurations \cite{algo}.}
The variance of the product of the $N$ factors is then reduced by a factor $1/N$.
In the limit $N\to\infty$ the exact determinant $\det(M^\dagger M)$ is obtained
and also the exact variance of the determinant due to the gauge fluctuations, i.e.
the variance in \eqref{eq:acceptance}. This provides
the possibility to analyse the exact acceptance rate as a function of the volume.

\vspace{-1em}
\section{Exact acceptance rate}
\label{sec:exact}
\vspace{-1em}

\begin{figure}[t]
 \centering
\begin{minipage}[t]{0.485\textwidth}
 \includegraphics[width=\textwidth]{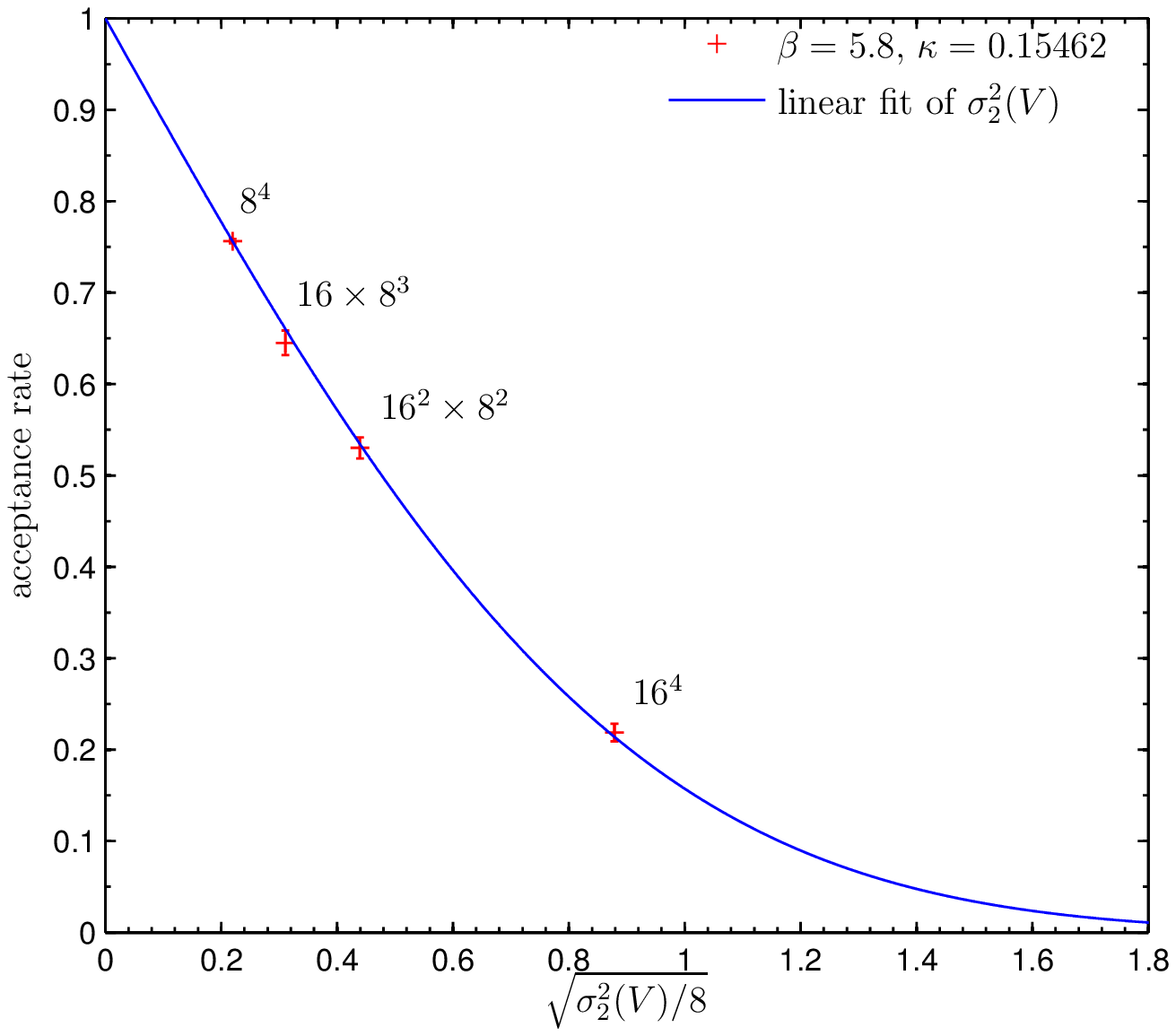}
 \caption[]{Exact acceptance rate for volumes (in lattice units) ranging from $8^4$ to $16^4$.
          A linear fit of the variance (inserted in \eqref{eq:acceptance}) describes the data well.}
 \label{fig:acc}
\end{minipage}
\hfill
\begin{minipage}[t]{0.485\textwidth}
 \includegraphics[width=0.99\textwidth]{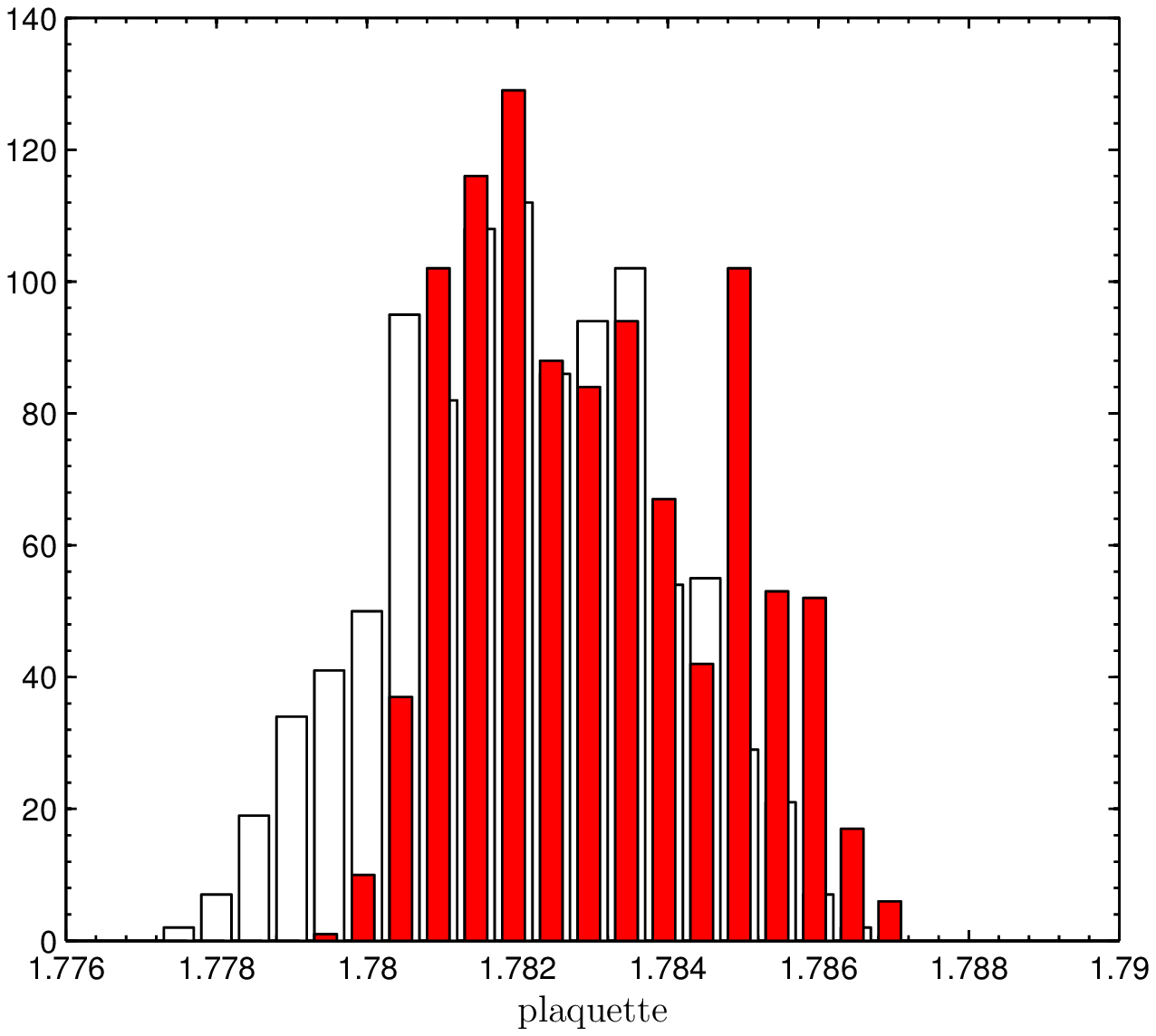}
 \caption[]{Histogram of the plaquette for the 3-step algorithm and a HMC run at the
            same parameters (cf.~section \ref{sec:run}) and using the same bins.}
 \label{fig:plaq}
\end{minipage}
\vspace{-1em}
\end{figure}

We consider a 3-step algorithm that first updates all active links of a domain decomposition
with block size $4^4$, which amounts to about 10\% of all links. This change is then
block-wise accepted with the exact determinant ratio of the block Dirac operators.
After some iterations of the combination of step one and two, the proposal is passed to the
global acceptance step with the Schur complement of the $4^4$ block decomposition.

The gauge action we use is the Wilson plaquette action and the Dirac operator is the
plain Wilson-Dirac operator. The lattice parameters are $\beta=5.8$ and $\kappa=0.15462$.
Theses values corresponds to a lattice spacing of $0.05\, \mathrm{fm}$ and a pion mass of
$400\, \mathrm{MeV}$ (both determined on a large lattice in \cite{DelDebbio:2005qa}). On
lattices with $8^4$ up to $16^4$ points we determine the variance of the stochastic estimator in the
global step for different values of $N$ in eq.~\eqref{eq:param} and extrapolate
in $1/N$ to zero, thus obtaining an estimate for the exact variance $\sigma^2_2(V)$ 
of the logarithm of the determinant as a function of the volume.
Via \eqref{eq:acceptance} the exact acceptance rate can also be determined from
the variance.\footnote{We tested the (tacitly assumed) validity of the Gaussian model for
finite values of $N$.}

The exact acceptance rates as determined from the variances are plotted in fig.~\ref{fig:acc}
together with the result of a linear fit to $\sigma^2_2(V)$ constrained to zero. The 3-step
algorithm of this section shows a good acceptance for lattices up to $16^2\times 8^2$.
This is the region where the error function can be approximated by a Taylor expansion
with a linear leading term. The data supports a linear growth of the fluctuations with the volume.

\vspace{-1em}
\section{Realistic run}
\label{sec:run}
\vspace{-1em}

The algorithm of the last section is characterised by the separation scale of
low and high modes given by the inverse block size $1/L_b$, with $L_b=4a$.
From fig.~\ref{fig:acc}
we conclude that this scale is too large for a lattice with the same parameters
and $16^4$ points. Indeed switching to a block size of $8^4$ and updating the 
links in a $6^4$ block inside each $8^4$ block (which amounts to 8\% of all links)
the global acceptance rate is raised to 61\% (at fixed $N=21$ and 45\% block acceptance).
A histogram of the
plaquette is shown in fig.~\ref{fig:plaq}. The plot comprises the data
of 1000 global acceptance steps after thermalisation. For comparison the histogram of the
plaquette obtained from a HMC run with the same parameters is plotted as well. The histograms
have to agree within statistical errors and we checked this for the mean value and the variance.

\vspace{-1em}
\section{Conclusion and outlook}
\vspace{-1em}

We have developed an algorithm that is based on a hierarchical filter of acceptance steps.
It deploys recursive domain decomposition to separate short and long distance physics.
On lattices with up to $16^4$ lattice points (or size $(0.8\; \mathrm{fm})^4$) the algorithm
has good global acceptance $\ge 50\%$.

We have already extended this study to larger volumes and other fermion actions including
the clover term and/or HYP-smeared links \cite{algo}.
The application of the techniques presented in sections \ref{sec:dd} and \ref{sec:stochastic}
to mass reweighting \cite{Hasenfratz:2008fg,Hasenfratz:2011} is promising and will be pursued.

%

\vspace{-0.75em}
\bibliography{text}
\bibliographystyle{JHEP}        

\end{document}